# Magnetization pinning in modulated nanowires: from topological protection to the "corkscrew" mechanism

*Jose A. Fernandez-Roldan\*, Rafael P. del Real, Cristina Bran, Manuel Vázquez and Oksana Chubykalo-Fesenko*

Instituto de Ciencia de Materiales de Madrid, CSIC, c/Sor Juana Inés de la Cruz, 3, 28049, Madrid, Spain

**Abstract**

Diameter-modulated nanowires offer an important paradigm to design the magnetization response of 3D magnetic nanostructures by engineering the domain wall pinning. With the aim to understand its nature and to control the process, we analyze the magnetization response in FeCo modulated polycrystalline two-segment nanowires varying the minor diameter. Our modelling indicates a very complex behavior with a strong dependence on the disorder distribution and an important role of topologically non-trivial magnetization structures. We demonstrate that modulated nanowires with a small diameter difference are characterized by an increased coercive field in comparison to the straight ones which is explained by a formation of topologically protected walls formed by two 3D skyrmions with opposite chiralities. For a large diameter difference we report the occurrence of a novel pinning type called here the "corkscrew": the magnetization of the large diameter segment forms a skyrmion tube with a core position in a helical modulation along the nanowire. This structure is pinned at the constriction and in order to penetrate the narrow segments the vortex/skyrmion core size should be reduced.

Among other possibilities, cylindrical magnetic nanowires could provide new opportunities for nanotechnological applications such as 3D magnetic recording (in which the data are stored vertically, for example, in the racetrack architecture[1]), actuators or sensors[2] and logical devices where the control over the position and motion of domain wall plays an essential role. It is well known that these nanowires are demagnetized via the domain wall propagation either of transverse or vortex type[3]. Importantly for applications, the domain wall velocity can achieve very high speed with the absence of the Walker breakdown[4].

The future of domain-wall logic devices, registers or race-track-memories requires precise control of domain walls dynamics including their pinning and depinning. Another possible application is the use of nanowires as permanent magnets where the engineering of the pinning of domain walls is necessary in order to increase the coercivity[5]. The domain wall pinning can be produced by artificial defects (e.g. notches[6]), including the nanowires with modulations in diameter [7-13] and multisegmented nanowires[14, 8]. Magnetometry measurements show a strong influence of the constrictions on the coercive field[7,10]. In addition, magnetic-force microscopy[13, 15, 10], electron holography[8, 11] and XMCD-PEEM imaging[15] indicate strong stray field coming from the diameter transitions where magnetic charges accumulate. They also have revealed the domain wall pinning at certain positions although the definite control and understanding of the pinning conditions are far from being unveiled.

In this article, we investigate theoretically the domain wall pinning in two-segment FeCo modulated nanowires. The FeCo nanowires are chosen due to their high magnetic moment[5, 16-17] which produces large longitudinal (shape) anisotropy with high coercive field[18]. The nanowires of this composition with uniform diameter have been reported to demagnetize via vortex (Bloch-point) or transverse domain wall propagation[3, 16]. However for the diameters of the present study only the vortex-type domain wall is present. The modulated FeCo nanowires are previously grown by electrodeposition inside the nanopores of anodic alumina templates[7, 17] and specially designed for periodical modulations in diameter. The electron holography imaging, at remanence, on FeCoCu modulated nanowires have shown the presence of the vortex states at the end of the large segments[11]. Importantly, measurements by Kerr-magnetometry on individual nanowires[17], when focusing on different spots along the wires, showed different local squared-shape hysteresis cycles with either a single or several jumps.

From the theoretical point of view, modulated nanowires composed by two segments with different diameters have been modelled considering Ni[19, 20] and FeNi[6,12,21,28]. The micromagnetic simulations show that the demagnetization process starts with the formation of the vortex domain wall in the segment of larger diameter. Under reverse field, these domain walls penetrate into the segment with smaller diameter and harder magnetic behavior[12, 21]. If this diameter is small and the difference between the two is moderate, then the transformation of the vortex domain wall into the transverse one could happen[19, 20]. No strong domain wall pinning is typically observed. In our case no conversion from the vortex to the transverse domain wall is expected due to higher saturation magnetization value. Besides, deterministic micromagnetic programs, used up to now, do not allow changes as, for example, of the vortex domain wall chirality. Thus, in our view, the existing theoretical studies are limited in materials, geometries, consider mostly two segments and do not clarify the reason of pinning observed in experiments with some exceptions.

Furthermore, the role of different non-trivial magnetization structure in hysteresis processes such as swirls, hedgehogs, helices have been discussed in the past[22, 23]. The helicoidal magnetization domain wall for example, have been introduced as possible magnetic configuration in iron whiskers[23]. However, a new burst of interest to topological structures has recently appeared in relation to the dynamics of vortices and skyrmions, topologically non-trivial magnetization objects, due to their very interesting dynamical behavior and potential applications as the information carriers[24]. Cylindrical magnetic nanowires offer an additional possibility to nucleate topological defects due to curling magnetization instabilities at the ends of the nanowires[25]. In this article, we show theoretically that magnetization processes in modulated magnetic nanowires are characterized by the existence of topologically non-trivial 3D magnetization configurations such as vortex and skyrmion tubes and their helicoidal structures. These structures are stabilized by the finite-size effects in nanowires of different diameters and play important role in the pinning process.

In order to elucidate the nature of the pinning we perform the micromagnetic modelling in the situation close to the experimental one. The nanowire geometry is presented in Fig. 1. We consider magnetic nanowires modulated in diameter with 5 segments and two different diameters (modulations). The two largest modulations have diameter D= 130 nm and length 1μm, and the narrow modulation has a variable diameter 40 nm < $d$ <100 nm and length 300nm. The constriction between the two diameters is 50 nm and the diameter is linearly varied from the smallest to the largest value. The two parts at the ends of nanowires have the minor and large diameters respectively and are intentionally considered to be different as depicted in Fig. 1 to induce an asymmetric propagation.

For the micromagnetic modelling of the demagnetization processes the state-of the art mumax3 code was used[26]. The micromagnetic parameters correspond to that of realistic experimental values[2, 3, 5, 16] with magnetization saturation $\mu_0 M_s = 2T$, the anisotropy energy constant $K_1 = 10^4$ $J/m^3$, and exchange stiffness $A_{ex} = 25\ pJ/m$. Nanowires are considered to have a granular bcc structure (modelled by the Voronoi tessellation with a grain size of 5nm) textured along [110] direction[16]. The other two easy axes components are assumed to be randomly distributed in the plane perpendicular to [110] direction in each grain[16,2]. Different realizations of disorder (i.e. the grain distributions) were modelled by changing the seed number for the random distribution.

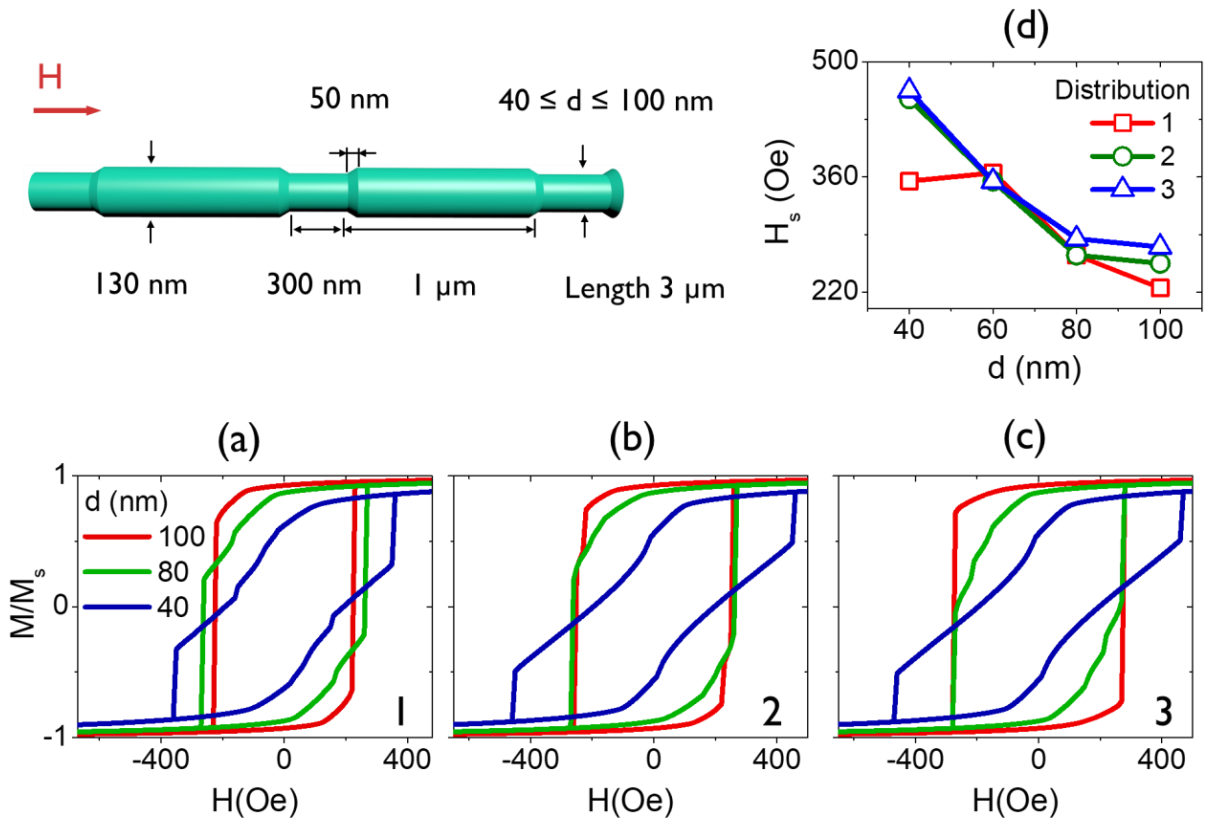

**Figure 1.** *(Top left)The geometry of the simulated nanowire. (a-c) Hysteresis loops obtained for three disorder distributions, labelled 1 - 3 respectively and for a minor constriction diameters 40, 80 and 100 nm. (d) Switching field as a function of minor diameter for the three distributions.*

Examples of hysteresis loops under field H applied parallel to the nanowire axis are presented in Fig. 1 for several minor diameters and grain distributions. Their shapes depend substantially on the disorder distribution, i.e. the particular polycrystalline microstructure of the nanowire. As for the common features, the hysteresis loops become less squared and more inclined in all cases as the minor diameter is reduced. The remanence decreases with the increase of the minor diameter and is practically independent on the considered distribution for each minor diameter d.

Every hysteresis loop exhibits a nucleation stage (the nucleation field increases with the increase of the minor diameter d), the magnetization propagation/rotation inside the segment of major diameter and the irreversible magnetization jump in the segments with minor diameter (depinning effect from constrictions). When the difference between diameters is small the propagation and the depinning part occur at the same field stage (this is called below "weak" pinning since the magnetization changes in large and narrow modulations occur at the same field stage but with an increased coercive field), see the case of d=100nm in Fig. 1. In nanowires with narrowest minor diameter (d<100 nm) the propagation of the vortices domain walls occur along the inclined slope before the magnetization jumps in small diameter modulations (" strong" pinning) as depicted in Fig. 1 by green and blue loops. In addition, the propagation stage is characterized by additional small jumps related to the change of internal structures in magnetization of the largest segments (see below). The field of the largest "jump" defines the switching field Hs. The value of $H_s$ typically decreases as a function of the difference between diameters, see Fig. 1(d). The particular local disorder as well as defects reduce or enhance the pinning field of the whole simulated nanowire as illustrated by diameters d=100nm and d=80nm in Fig. 1(c). However, generally the simulations show a high pinning field ("strong" pinning) for the case of the large difference between the constrictions diameters. Some rare exceptions of the general tendency also depend on particular disorder distribution as evidenced for the distribution 1 and d=40 nm corresponding to Fig. 1(a).

The detailed analysis of magnetization configurations during the hysteresis processes shows that each case has its own peculiarities. It is well known that the switching field is smaller for nanowires with larger diameter[3]. Consequently, we observe that the demagnetization process always starts in the segments of larger diameter (or at the nanowire ends). This is also in agreement with previous micromagnetic modelling results on Ni and FeNi in two-segments[6,19-21]. As a common observation, the hysteresis process always starts with the formation of the vortex structures at the nanowires ends, see Figs 2(a, b). In the completely ideal case the vortices should appear in the alternating chirality pattern due to the dipolar energy minimization. However, the state with the same vortex chiralities in the nanowire of 1 µm (as the length of our large segments) has a very similar energy. Importantly, the granular structure has a major role here promoting a completely random chirality pattern, see Fig. 2(a-b) and the analysis in the Supplementary Information (SI). The formation of the vortex structures is followed by further rotation of magnetic moments inside the larger modulations leading to the formation of vortex tubes which expand from the ends towards the center of the modulation. As the field becomes more and more negative, the spins in the outer shells rotate towards its directions and the vortex tubes propagate inside the segments of larger diameter. Consequently, they become the Bloch skyrmion tubes (see Fig. 2(c-d) as a particular example) since the spins in the outer shells are directed antiparallel to the core direction. These skyrmions are of the Bloch (bubble) type, i.e. they have an arbitrary chirality since they are stabilized by the nanowire border and not by the Dzyaloshinkii-Moriya interactions as in chiral magnets[27]. However, they have the same topology and bear a non-zero topological charge.

The penetration of these structures into the constrictions with small diameters corresponds to the switching field (different in each case) and takes place in one step if the difference between the diameters is small (see SI for more details). The switching field depends on the particular chirality pattern, if the vortices/skyrmions in the same modulation are formed with the opposite chirality, when the two tubes with opposite chiralities meet, the resulting complex domain wall (called helical domain wall in Ref.[28]) has a strong topological protection and is difficult to annihilate. Particularly, the reversal of magnetization in smaller segment may take place before the complete switching in the segment of larger diameter. For a detailed analysis of the coercivity for each disorder distribution see the SI.

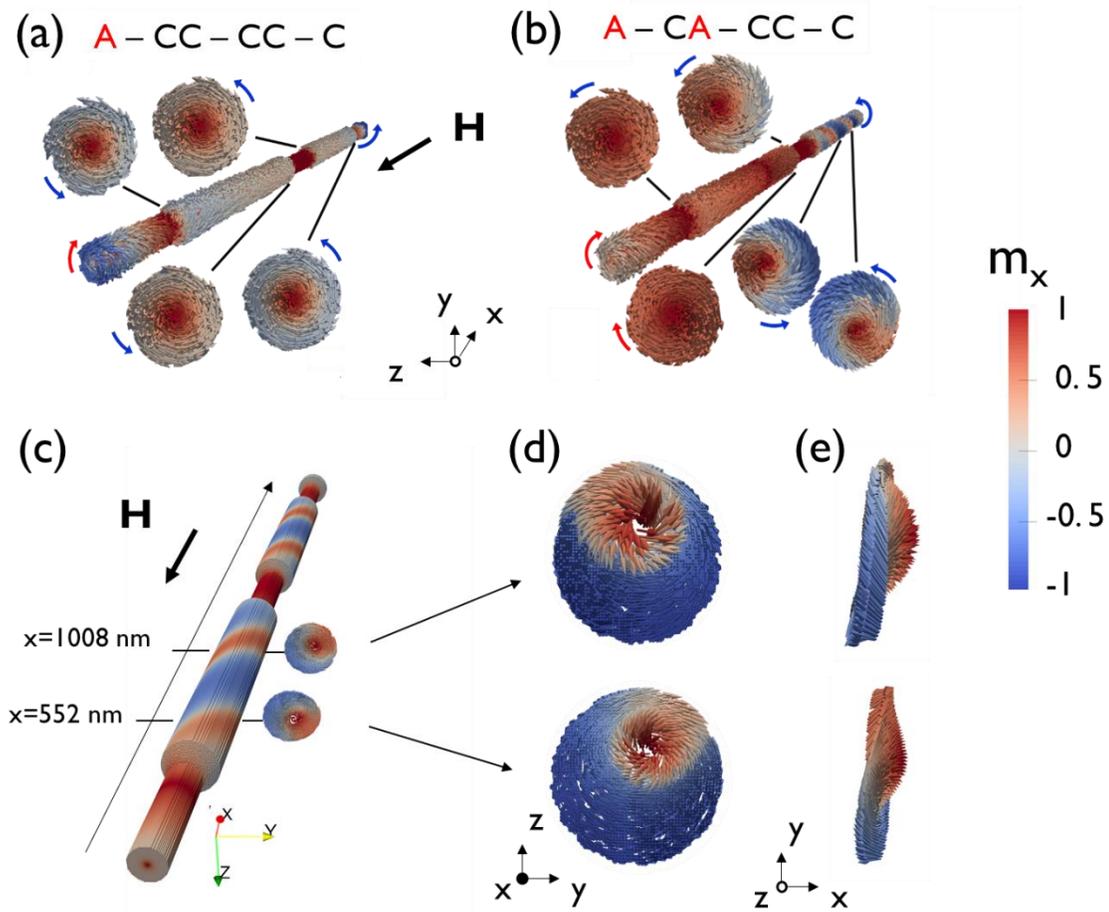

**Figure 2.** *Simulated magnetization distribution in two-segment nanowires with a particular disorder and small difference between diameters just before the switching field. (a-b) Longitudinal magnetization component at the nanowire surface and in the cross sections for d=100 nm for grain distributions with hysteresis cycles of Fig. 1(a, b) respectively ($m_x$>0 red color, $m_x$<0 blue color and grey color for $m_x$ close to 0. C stands for the clockwise and A for the anticlockwise chiralities of the vortices along the length of each nanowire. (c) Surface magnetization distribution in a nanowire with disorder distribution No.1 with d=80 nm just before the switching showing the magnetization spiral and the vortex/skyrmion magnetization structure (d-e) views of cross sections of the nanowire at two positions showing the displaced skyrmion structure. The right color map shows color scale for the whole figure.*

In the opposite case of large diameter difference, we have a "strong pinning", i.e. the skyrmion tubes structures become pinned at the constrictions and an additional field is necessary to unpin them. Again, the chiralities are formed with an arbitrary pattern and the vortices are transformed into the skyrmion tubes as the applied field becomes negative. For the detailed analysis of particular hysteresis process and the influence of the chirality pattern see the SI. In order to understand the nature of the pinning, in Fig. 3(a) (bottom panels) we indicate by the red color the regions where the magnetization is directed along the nanowire before the magnetization unpinning (locus of the magnetization with $m_x>0.97$). The narrow segments (seen here as wide) are still magnetized parallel to the core. Along the wide segments skyrmion tubes (with the shell magnetized parallel to the field and the core - antiparallel to it) are present, see also Fig. 4(b). In the bottom panels of Fig. 3(a) one can clearly see the oscillatory behavior of the skyrmion core position for large diameter differences. This can be also seen from the presentation of the magnetization components along the nanowire center in Fig. 3(c). The skyrmion core position thereby forms a helicoidal structure, pinned at the constrictions between segments, called here the "corkscrew" pinning.

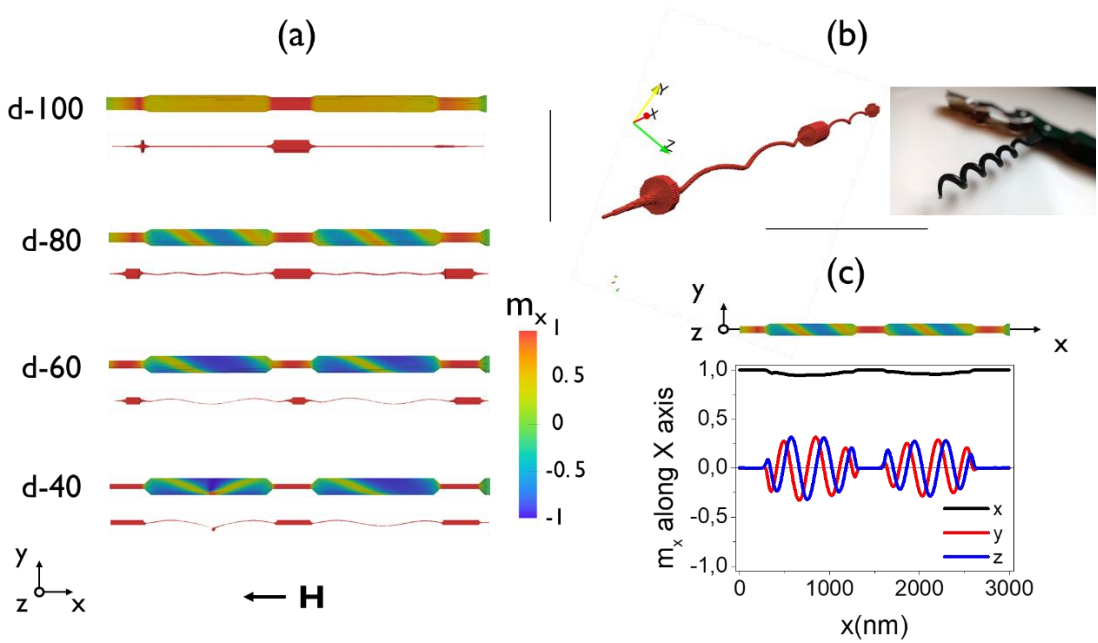

**Figure 3.** *(a) The longitudinal magnetization component on the nanowire surface (upper images), and (bottom images) the magnetization with the locus of $m_x>0.97$ for each minor diameter for the distribution No. 1. (b) The locus of $m_x>0.97$ in the nanowire with minor diameter 60 nm (left) is compared with the helix of a corkscrew (right). (c) Magnetization components in the nanowire center along the x axis in the nanowire with minor diameter 80 nm.*

To further understand its nature, we analyze the vortex (d=100nm)/skyrmion (d=40-80nm) shape before the switching at a constant distance from the constriction, see Figs. 4(a-c). Following the helicoidal structure, its center is displaced from the nanowire center and the helicoidal amplitude increases with the increase of the difference between diameters. Note that for this particular disorder distribution the vortex for d=100nm is placed in the nanowire center, however, depending on the disorder we also found a helicoidal behavior for this diameter, see SI. Fig. 4(c) presents the vortex/skyrmion profile along the direction of the line joining its core and the nanowire centers as depicted by red dark arrows in Fig. 4(b). The displacement of the skyrmion core from the nanowire center is larger for a larger difference of the diameters indicating that in order to penetrate into the small modulation the core is first displaced (in other words, repelled from the constriction to minimize the magnetostatic charges). The magnetic charges are then redistributed over the whole volume forming a spiral. The spiral amplitude and the frequency increases with the diameter difference. Another effect is the narrowing of the skyrmion core in order to get accommodated into the narrower modulation. Since the skyrmion core size depends on the applied field, a larger field is necessary when the difference between the diameters is larger.

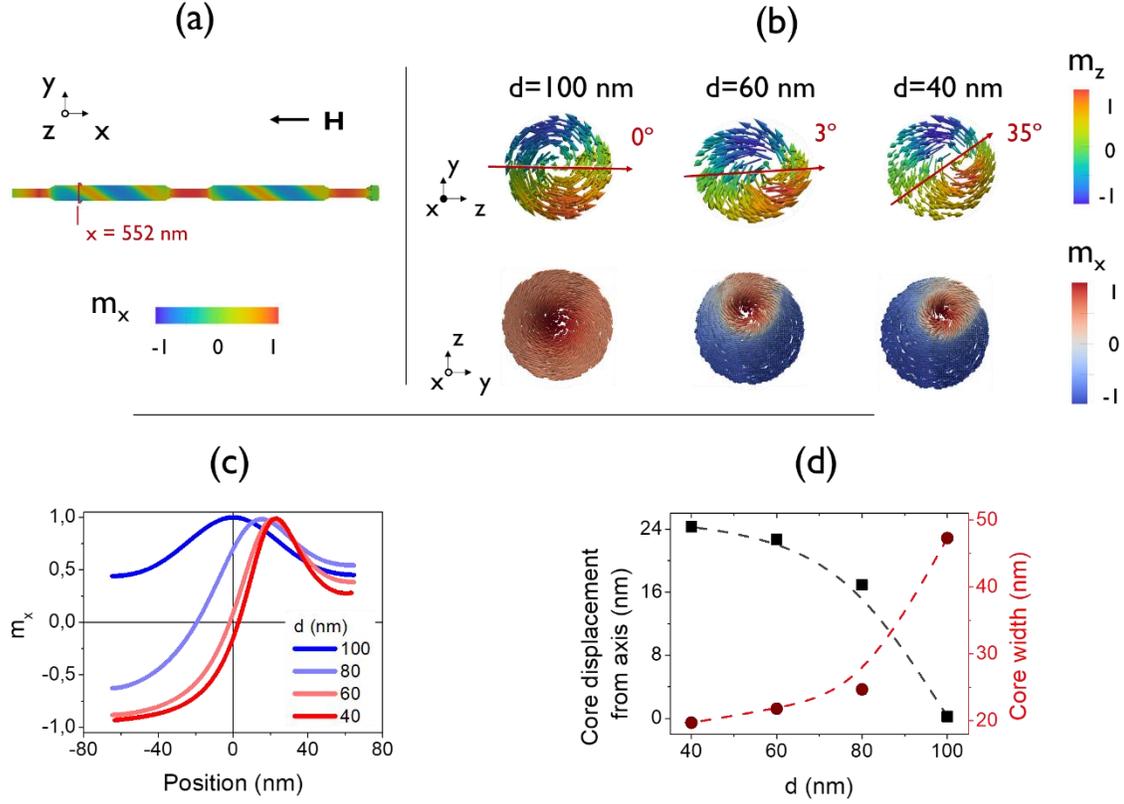

**Figure 4.** *(a) A nanowire and its longitudinal magnetization distribution showing the fixed position used for the analysis of the vortex/skyrmion for each nanowire. (b) Cross sections of the magnetization at the marked position for three different nanowires of the disorder distribution No. 1 colored by $m_z$ (top) and $m_x$ (bottom). The red arrows join the nanowire and vortex centers for each cross section and indicate the direction along which the core size was measured. (c) Longitudinal component of the magnetization along the red arrows for the nanowires of distribution No.1. (d) Core width and displacement from the nanowire axis as a function of the minor diameter.*

In conclusion, we have shown that the modulated nanowires can be engineered for magnetization pinning by varying the diameter of the narrow segment. The magnetization process occurs by vortex formation in large segments with arbitrary chiralities. As the field increases in the negative direction, the vortices expand in length and form tubes which are later transformed into the skyrmion tubes. Thus, the presence of topologically non-trivial configurations is inherent for the magnetization reversal processes and defines the pinning nature. As a general rule, the larger the difference between diameters, the stronger the pinning.

The value of the pinning field is significantly affected by two main contributions: (i) the difference in diameters between the constrictions, and (ii) the particular disorder which determines the chiralities of the vortex domain walls.

For nanowires with moderate difference between the two segments diameter, the switching field is determined by the annihilation of skyrmion tube structures. When two of these tubes with the opposite chiralities meet, the resulting structure is strongly topologically protected and requires and additional field. The uncontrollable vortex chirality suggests the difficulty of precise control of some properties such as the coercive or switching field of modulated nanowires.

For larger difference between diameters, a strong pinning is observed and a much stronger field is required to switch the small diameter modulations. The corresponding constrictions are characterized by stronger magnetostatic charges and, in order to minimize them, the skyrmion core is displaced from the nanowire axis. Here we report a new type of pinning, called corkscrew-like, since the skyrmion core positions form a helicoidal structure. The amplitude of this structure increases as a function of the diameter difference and the skyrmion core size near the switching field decreases. Our analysis introduces a new type of pinning and offers novel perspectives to its control. This is important for geometrical design of magnetic nanowires aiming at multiple nanotechnological applications.

**Associated content**

The Supporting Information is supplied in an additional PDF. This includes further analysis, comments and figures.

AUTHOR INFORMATION

**Corresponding Author**

* José A. Fernández-Roldán. jangel.fernandez.roldan@csic.es

**Author Contributions**

M.V., R.P., C.B. and O.C.-F. have conceived the idea of the manuscript. J.A.F.-R. has conducted the micromagnetic simulations and postprocessing with the assistance of O.C.-F., R.P., and M.V. The results were analyzed and discussed by all authors during the whole process. The main text was written by J.A.F.-R. and O.C.-F. All authors have written, reviewed and approved the final version of the manuscript.

**Notes**

The authors declare no competing financial interest.

**Acknowledgment**

This work was supported by the Spanish Ministry of Economy, Industry and Competitiveness (MINECO) under the grants MAT2016-76824-C3-1-R and FIS 2016 – 78591-C3-3-R. J.A. F.-R. acknowledges support from MINECO and ESF though fellowship BES-2014-068789.

# Magnetization pinning in modulated nanowires: from topological protection to the "corkscrew" mechanism

*Jose A. Fernández-Roldán\*, Rafael P. del Real, Cristina Bran, Manuel Vázquez and Oksana Chubykalo-Fesenko*

# Supplementary information

**S1: Magnetization configurations during the magnetization reversal processes**

We present a more detailed analysis of the configurations during the magnetization reversal processes. For a small diameter difference, this happens in one field step so that a dynamical magnetization evolution is presented in Fig. SI1. The vortices are formed with the chirality pattern A-CC-CC-A. The vortices centers also form a helicoidal structure in this case. The reversal starts with the vortex cores reduction, the skyrmion tubes formation and propagation inside the middle narrow segment (Fig. SI1a-c) where the size of the core has been largely reduced. The tube breaks into two tubes by the formation of Bloch points (Fig. SI1d). The Bloch points propagate with different velocities in opposite directions (Fig. SI1d-g). During this propagation, the structures formed by the skyrmion tubes with opposite chiralities at the first left and last right modulation transitions are annihilated simultaneously due to the high concentration of magnetostatic charges in those regions as represented in Fig. SI1e. Simultaneously several new Bloch points are formed and start propagating in opposite directions (Fig. SI1e-g). The effect of different chiralities can be observed in Figs. SI1 e and g: we see that the magnetization reversal is the last to happen in the left and right ends where the vortex chiralities are opposite.

The reversal process is retarded there due to topological protection between opposite chiral skyrmion tubes, which is absent in the middle section.

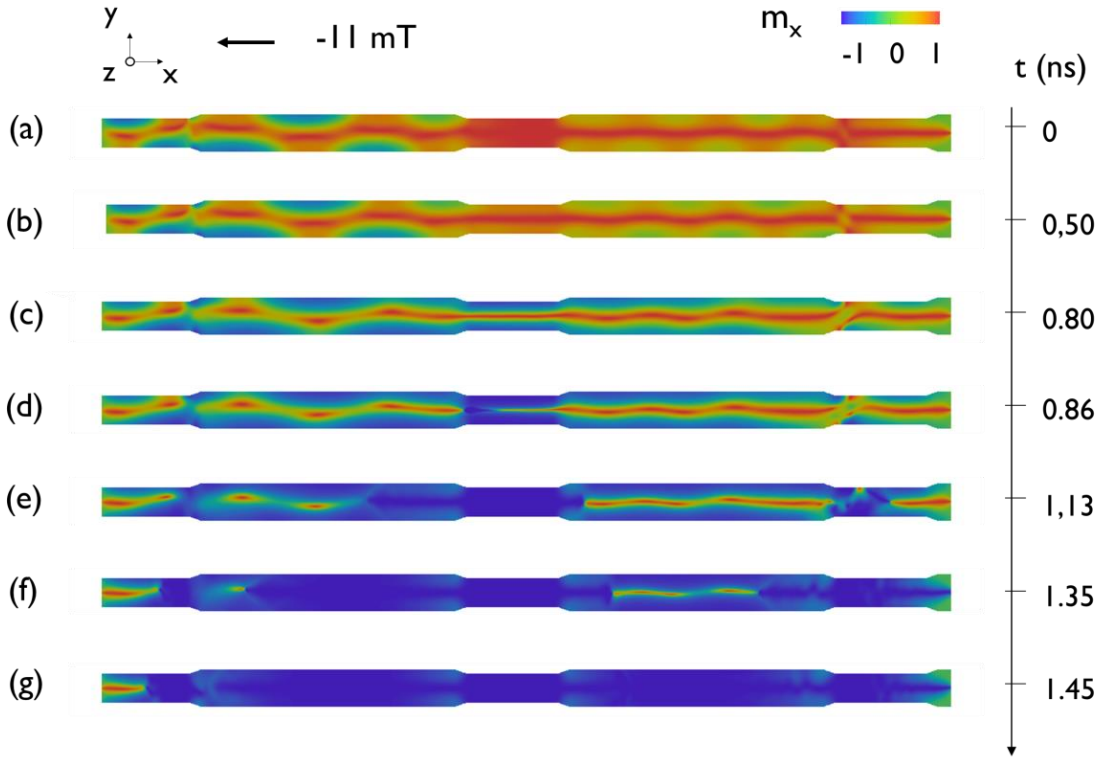

**Figure SI1.** *(a-g) Longitudinal component of the magnetization in the middle cross section of a nanowire with diameter 100 nm and disorder distribution No.1 at different times during reversal. The nanowire exhibits vortices with chiralities A-CC-CC-A along the nanowire length.*

An example of the reversal process is presented in Fig. SI2 for nanowires with the minor diameter 40 nm and the disorder distribution No.1 (Fig. 1a). The nucleated vortices at the modulation and at the nanowire ends quickly propagate inside the larger modulations and as the field is gradually reduced below zero they become skyrmion tubes. In the left large segment, the vortices have opposite chiralities at the two ends while in the right one show the same chiralities.

Consequently, the propagation in the right segment is easy and at H=-170 Oe only one skyrmion tube with a unique chirality is present (See Figure SI2a). In the left wide modulation the vortices at the end of skyrmion tubes with the opposite chirality cannot be easily annihilated and when met, they are divided by a complex domain wall[25] (called a helical domain wall in Ref. 28). Additional field is thus required to annihilate this structure. At H=-180 Oe a small magnetization jump appears in the hysteresis loop (Fig. SI2a), related to a transformation of this wall to a different configuration characterized by a vortex and an antivortex on the surface of the wire (Fig. SI2e). This new structure becomes larger as the field is further increased (See vortex in the inset picture in Figure SI2d). Overall, the narrow segments remain uniformly magnetized until H=-455 Oe when the complete switching takes place in one irreversible jump.

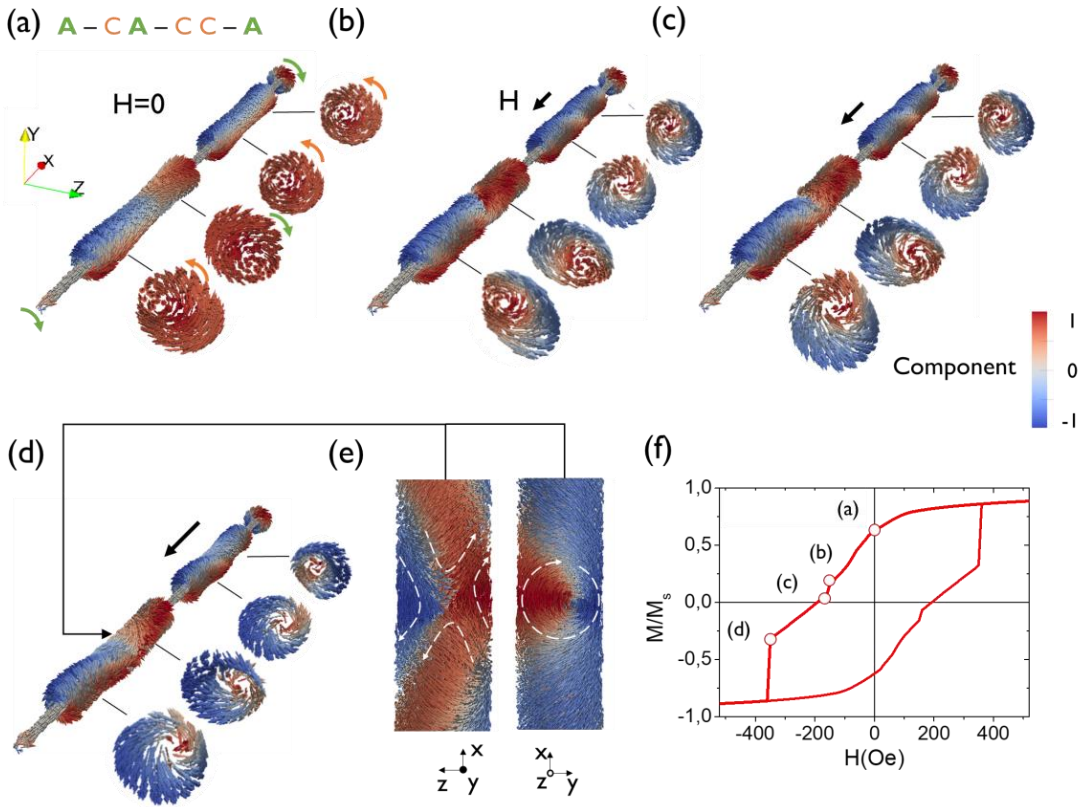

***Figure SI2.*** *Longitudinal component of the surface magnetization of a modulated nanowire with a minor diameter of 40 nm for distribution 2 at different fields marked in the hysteresis cycle (f). (a) remanence, (b) -170 Oe, (c) 180 Oe and (d) and (e) -450 Oe before and after the jump. Inset figures are transverse cross sections of the nanowire at the marked positions in (b-d) where the colors indicate z-component. (e) A cropped perspective of the segment surface longitudinal magnetization at the marked site. Red and blue arrows show the chirality of each vortex/skyrmion. C and A stand for the, Clockwise or Anticlockwise chirality, respectively.*

### S2: Analysis of the chirality influence

The magnetization switching in these nanowires takes place by the formation of vortex domains with arbitrary chiralities in the larger segments which should penetrate into the smaller ones. It is clear that there are many possibilities of different chiralities patterns. For small difference in segment diameter we present below the transverse magnetization components and the chirality pattern for the three disorder distributions with hysteresis cycles presented in Fig. 1. On clearly see the arbitrary chirality pattern which is summarized in Table SI1. The patterns are labelled as X-XX-XX-X, with X, either A or C for Anticlockwise of Clockwise vortex domain chirality, respectively.

|  | d (nm) | 100 | 80 | 60 | 40 |
|---|---|---|---|---|---|
| Distribution | 1 | A-CC-CC-C | A-CC-CC-C | A-CC-CC-C | A-CA-CC-A |
|  | 2 | A-CA-CC-C | A-CA-CC-C | A-CA-CC-C | A-CC-CC-C |
|  | 3 | A-CA-CA-C | A-AA-CC-A | C-AA-AC-A | A-AA-AA-A |

***Table SI1.*** *Chiralities of the vortex structures nucleated at the ends of the wire and at the ends of modulations for each distribution and minor diameter. C (A) indicates the Clockwise (Anticlockwise) chirality following a scheme X-XX-XX-X along the nanowire profile according to Fig. 1.*

By examining the magnetization structures during the hysteresis process in each case we can conclude that in the case of the large difference between the segment diameters, the chiralities of the formed vortex domains have a small effect in the depinning field and the disorder seems to play the crucial role. Consider as an example the disorder distribution corresponding to the case No.3 and d=40nm (Fig. 1c, "strong" pinning) which produced a vortex pattern with the same chirality along the whole nanowire and have the largest depinning field. However, the chiralities of different vortices have a large effect during the propagation stage (which affects the coercive field) resulting (or not) in additional jumps corresponding to the annihilation of vortices of different (or the same) chirality but not on the depinning field itself.

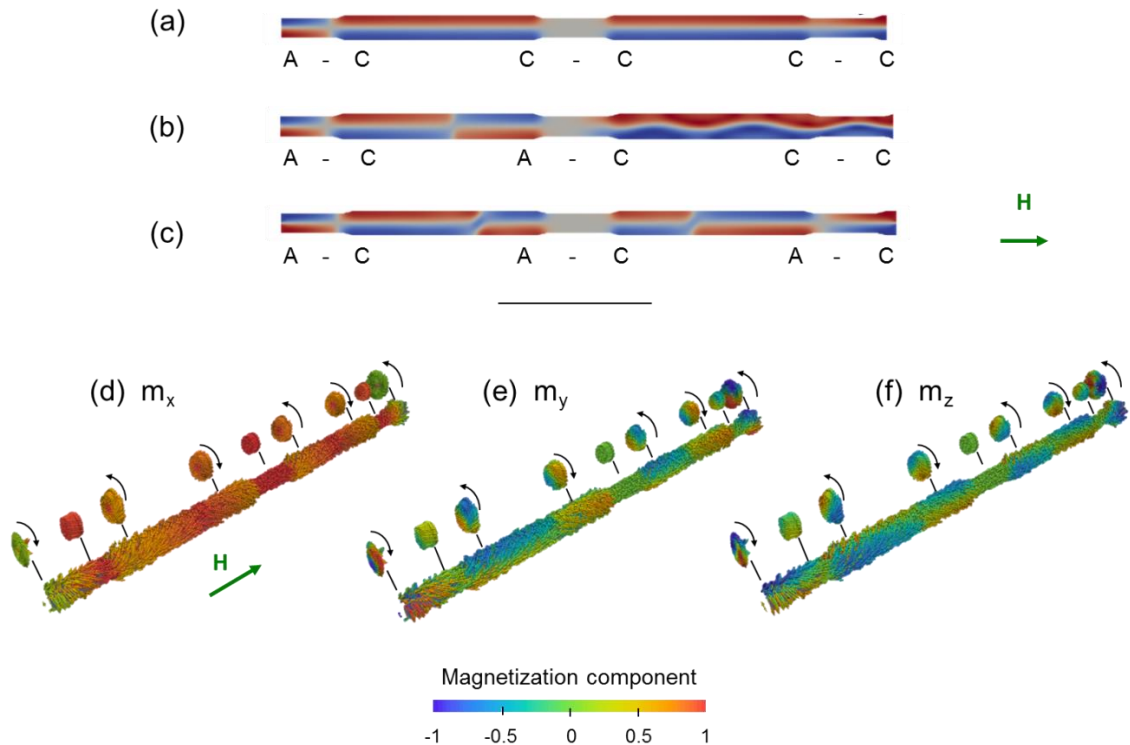

*Figure SI3*. *(a-c) Schematic representation of the transverse magnetization component inside the nanowires in the middle cross section along the nanowire profile before the switching for d=100 nm for*

*grain distributions with hysteresis cycles of Fig. 1 ($m_y>0$ red color, $m_y<0$ blue color and grey color for $m_y$ close to 0). Fig.(a-c) correspond to the grain size distributions labelled (1-3) of Fig. 1, respectively. The chirality of each vortex is shown by C(Clockwise) or A (Anticlockwise). (d-f) Surface magnetization distribution corresponding to the nanowire depicted in Fig. SI3(c) showing also the cross-sections with the vortex chiralities.*

The conclusions are different when the difference between the diameters is small ("weak" pinning). In this case the observed chirality pattern plays a major role since the propagation and depinning stage occurs here at the same field. Let us analyze in detail the case with a minor diameter d=100nm. Fig. SI3(a-c) shows schematically the patterns observed for this case. The lowest $H_s$ =-225 Oe corresponds to the nanowire with the distribution 1, for which all vortices have the same diameter and the nanowire nucleates in an almost uniform pattern: clockwise vortices at the ends of all modulations and at the right end of the wire, and a small anticlockwise vortex part on the left end of the wire, see Figure SI3 (a). These domain walls easily propagate along the largest segments and the switching field is minimum. Nevertheless, for disorders 2 and 3, different chiralities are found for the nucleated vortex structures, which lead to different switching fields: H=-255 Oe (for distribution 2) and H=-275 Oe (for distribution 3). When meeting inside a large segment, the vortex domain wall (which have become the skyrmion tubes) with different chiralities produce a complex 180-degree domain wall which requires an extra Zeeman energy for annihilating. The highest switching field value for d=100 nm is obtained for the distribution 3. In this case a completely alternating pattern of vortices with opposite chiralities is produced (Figure SI3 (c)). Note also the helicoidal structure for the vortex domain for the disorder No.2

### S3: Analysis of the helicoidal structure

The helicoidal structures presented in Fig. 3(a-b) have been analyzed for the other distributions before the switching field. The results are summarized in Fig SI4 (d-e). The typical "corkscrew" helicoidal shape for the core positions of the skyrmion tubes is depicted in SI4(a-b) corresponding to different selected diameters. They are characterized by the uniform magnetization at the minor segments and helicoidal tubes connecting them. The effect of the minor diameter variation is the decrease of the amplitude of the helix with d, which are less frequent for d=100nm SI4(c), but not completely absent for every distribution SI4(d). The helix chirality (the handedness of the helix) is found to be arbitrary and the pitch is not uniform along the large segments, becoming larger in the middle of the large segments in the absence of tubes of opposite chiralities.  It is also noticed that the minor segments are almost magnetized uniformly for larger diameters which can be seen comparing the last right minor segment for each diameter in Fig SI4(d). The minimization of magnetic charges at the diameter transition regions leads to partial demagnetization of the minor segments. It also makes the skyrmion tubes longer by allowing them to partially penetrate into the minor segments. This reduction of magnetized area in the minor segments due to magnetostatic energy minimization increases the switching field observed in Fig. 1.

As is previously mentioned, the effect of the opposite chiralities of two consequent vortices is to create a topologically protected structure which may be present for every diameter even without the corkscrew shape. These structures are seen in Figs. SI4(b, c). This effect is even stronger for minor diameters and leads to the high switching fields in Fig. 1 as a consequence of the confinement of the skyrmion tubes inside the minor segments.

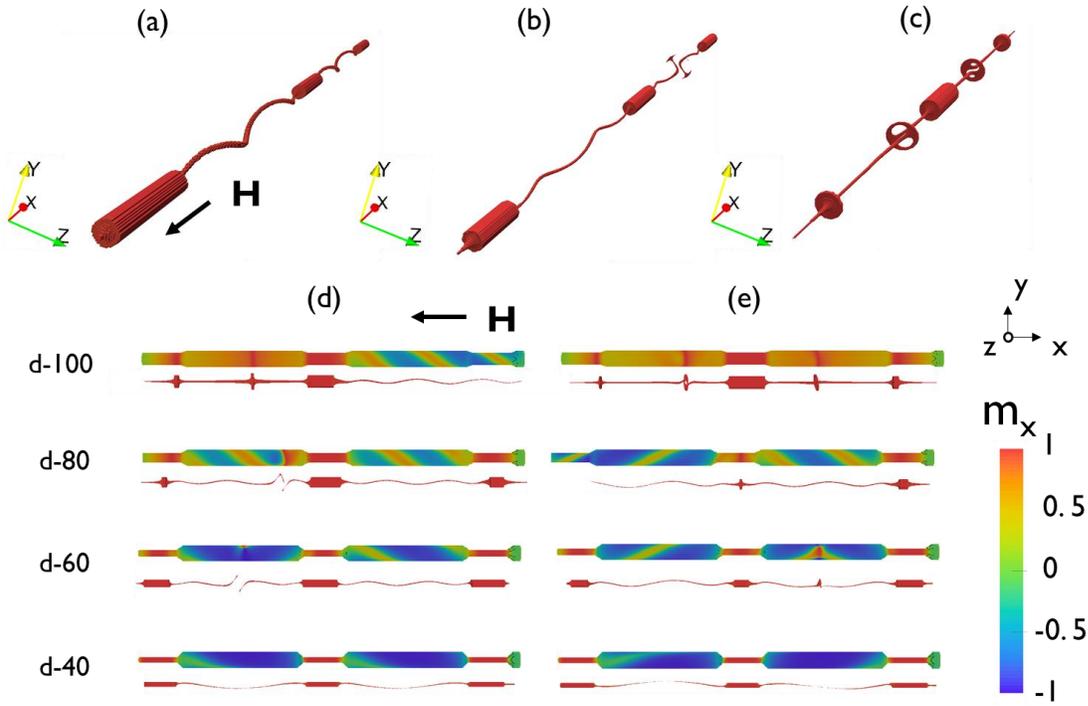

***Figure SI4.*** *(a) Locus of $m_x>0.97$ in the nanowire with minor diameter 40 nm of distribution No. 2. (b) and (c) locus of $m_x>0.97$ in the nanowires with minor diameters 60 and 100 nm of distribution No.3. (e-f) From top to bottom, the longitudinal component of magnetization on the surface, and the locus of $m_x>0.97$ for each minor diameter of the nanowires of distribution No. 1*

Fig. SI4(d-e) show that the information of the inner part and the surface magnetization in a nanowire are correlated. Thus, measuring the longitudinal magnetization over the surface of the nanowire can give an idea about the magnetization inside. Summarizing, the following information form the inner part is encoded on the surface: First the presence of vortices/skyrmion tubes by a color magnetization gradient. If those tubes have opposite chiralities, the position where they meet can be inferred on the surface as an abrupt area where the magnetization is not reversed as a result of the vortex core displacement to the shell as is seen for diameters 80 and 60 nm from SI4(d) and 60 nm in SI4(e). Furthermore, the presence of a corkscrew structure is characterized by a twisted pattern of two opposite magnetization values on the surface and presents the same pitch as the inner helix. The chirality of the helix is also defined by the twisting direction. On the other hand, the transverse component of magnetization leads to a direct measurement of the chiralities of the vortices as shown in Fig SI3. In the case of 100 nm for SI4(d) helix also propagates inside the minor information. The latter provides useful information for analyzing XMCD-PEEM experimental data from the surface and the inner parts.

The characteristics of the core of the skyrmion tube have been studied at the fixed position in Fig 4a) along one line joining the nanowire center and the core of each skyrmion for each wire as shown in Fig 4 (b) and Fig SI5(a-b). The rotated angle of the core along the length is arbitrary. The longitudinal magnetization component along that line shows that the core of the skyrmion tube is magnetized along the saturation field direction (Fig. SI5(c-f)). The core width is reduced for narrow minor diameters and displaced form the center of the nanowire. The core width and displacement values are independent of the distribution with an eventual exception ascribed to the particular disorder differences.

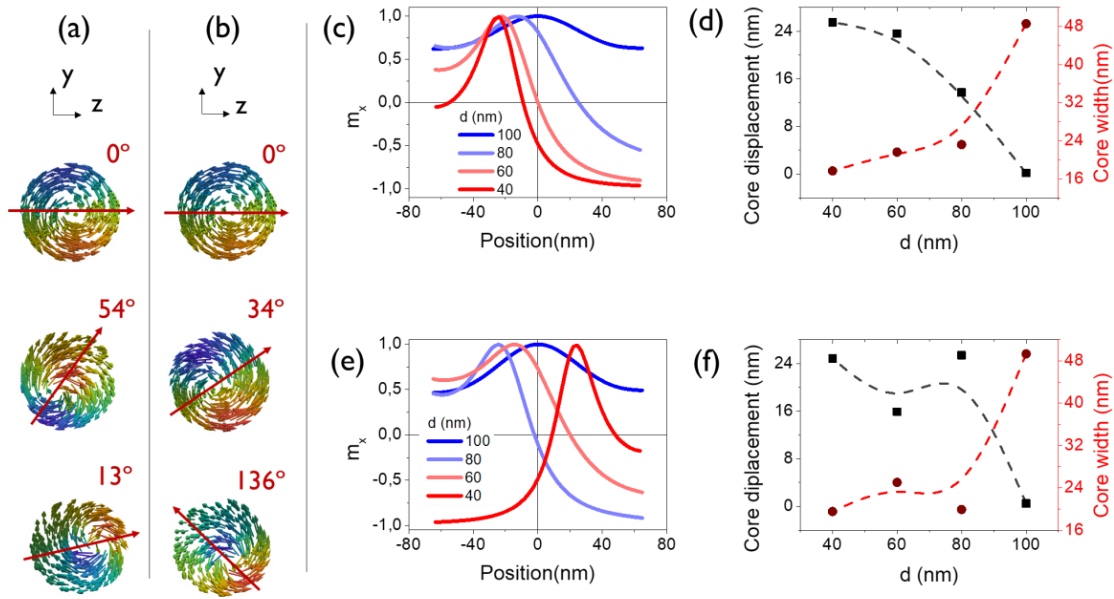

*Figure SI5.* *(a-b) Cross sections of the magnetization at the marked position in Fig 4(a) for nanowires with minor diameters 100, 80 and 40 nm (top to bottom) of distributions No. 2 and 3 respectively. Red arrows join the nanowire and the vortex/skyrmion centers for each cross section. (c) and (e) Longitudinal*

*component of the magnetization along the red arrows for the nanowires of disorder distributions No.2 and 3, respectively. (d) and (f) Core width and displacement from the nanowire axis as a function of the minor diameter for disorder distributions No. 2 and 3.*

The longitudinal magnetization components along the nanowire axis have been investigated for each distribution and diameter and confirm the non-periodic pitch of the helix in Fig. 3(c) and FigSI6(a-f). The presence of the corkscrew is determined by a large drop of the longitudinal magnetization component which has a "valley" -like shape when the initial vortices have the same chirality FigSI6(b-c, e-f) and presents a peak when they have opposite chiralities. The partial demagnetization of the minor segments is characterized by a step at each end of the catenary curve. For the nanowire with d=100nm of the disorder distribution 2, the vortex is deformed by the penetration inside of the minor diameter and there is a large shift of half catenary to lower values. The other magnetization components show information about the helical curling of the skyrmion tube. Despite the lack of periodicity of the helical structure, a quasiperiodic behavior is observed in the cases of vortices with same chirality in the modulations (Fig. SI6 (a-c, e, f)) which are particularly clear in Fig. SI6 (c, f) and Fig. 3(c): The local maxima of the oscillations of y and z magnetization components in the first modulation of Fig. 3(c) are separated by 308, 368 and again 308 nm, while in the second modulation by 294, 351 and 274 nm.

The complexity and the rich diversity of situations, begin a consequence of the particular disorder distribution of each nanowire, motivates further studies on the influence of the chiralities patterns in modulated nanowires for the future advanced technological applications.

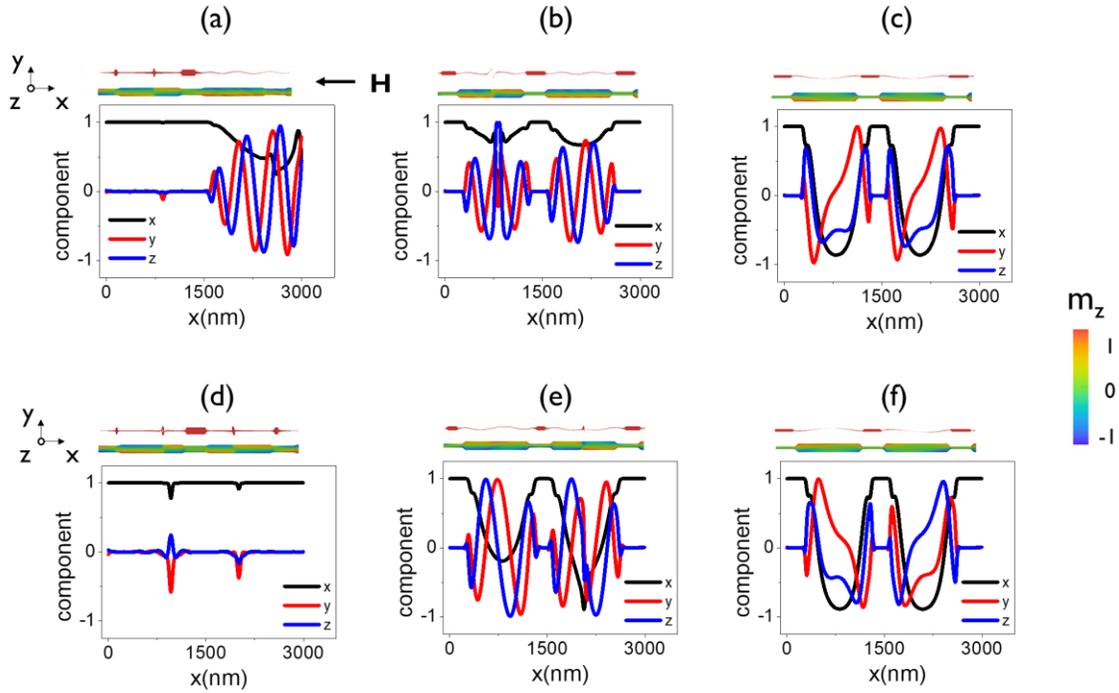

***Figure SI6.*** *(a) and (b) Longitudinal magnetization components along the nanowire in nanowires with minor diameters d=100, 60 and 40 nm for disorder distributions No. 2 and 3, respectively. In each graph the geometry of the nanowire (bottom) and the locus of magnetization with $m_x > 0.95$ (top) are shown.*